\documentclass[runningheads]{llncs}
\usepackage{graphicx}
\usepackage{float}
\usepackage{subfigure} 
\usepackage{amsmath,amssymb} 

\usepackage{color}

\begin{document}
\pagestyle{headings}
\mainmatter

\title{Real Image Restoration via Structure-preserving Complementarity Attention} 
\titlerunning{ }
\authorrunning{ }

\author{\bf{Yuanfan Zhang, Gen Li \thanks{Corresponding author.}, Lei Sun}}
\institute{Platform Technologies, Tencent PCG\\
\textit{{jklovezhan}@tencent.com}; \textit{{leegeun@yonsei.ac.kr}}
\textit{{raylsun}@tencent.com}}
\maketitle

\begin{abstract}
Since convolutional neural networks perform well in learning generalizable image priors from large-scale data, these models have been widely used in image denoising tasks. However, the computational complexity increases dramatically as well on complex model. In this paper, We propose a novel lightweight Complementary Attention Module, which includes a density module and a sparse module, which can cooperatively mine dense and sparse features for feature complementary learning to build an efficient lightweight architecture.
Moreover, to reduce the loss of details caused by denoising, this paper constructs a gradient-based structure-preserving branch. We utilize gradient-based branches to obtain additional structural priors for denoising, and make the model pay more attention to image geometric details through gradient loss optimization.Based on the above, we propose an efficiently Unet structured network  with dual branch, the visual results show that can effectively preserve the structural details of the original image, we evaluate benchmarks including SIDD and DND, where SCANet achieves state-of-the-art performance in PSNR and SSIM while significantly reducing computational cost.
\dots
\end{abstract}

\section{Introduction}
Image denoising is a task of restoring high-quality images from its noisy inputs. Due to the ill-posed nature, this is a very challenging task, which usually requires a strong image prior to be effectively restored. Because convolutional neural networks (CNN) perform well in learning generalizable priors from a large-scale data, they have become the first choice Compared with traditional image algorithm. However, although numerous CNN based approaches have been achieved impressive results, few methods simultaneously take care of aggregation features in terms of both dense and sparse module to more conducive to preserving the structural details. To this end, in this paper, we aim to save structural  details by exploiting the dual branch Unet architecture through a hierarchical complementary attention mechanism and a gradient structure maps.

Therefore, we propose a novel efficient yet effective SCANet architecture for image denoising. The proposed SCANet adopts a pixel-based branch with Unet structure \cite{ronneberger2015u} and a gradient-based branch via residual skip connections.

The proposed SCANet has three core designs that makes it suitable for image denoising. The first core design is the Complementary Attention Module(CAM), which consists of both dense and sparse module. For the dense module, the integrated features are extracted through the both spatial attention and channel attention; For the sparse module, it replaces the traditional convolution with group convolution and linear cheap computation, which greatly reduces the computational complexity while allowing the model to focus on sparse features, and the complementarity of the two brings powerful feature representation capabilities to the model. The second is to construct additional gradient based branches, which can retain more structural details by introducing gradient structure information. The third is the overall architecture adopts the hierarchical encoder-decoder unet structured design, which can efficiently and simultaneously extract the informative features with multiple receptive fields.

\begin{figure}[htbp]
    \centering
    \includegraphics[width=12cm]{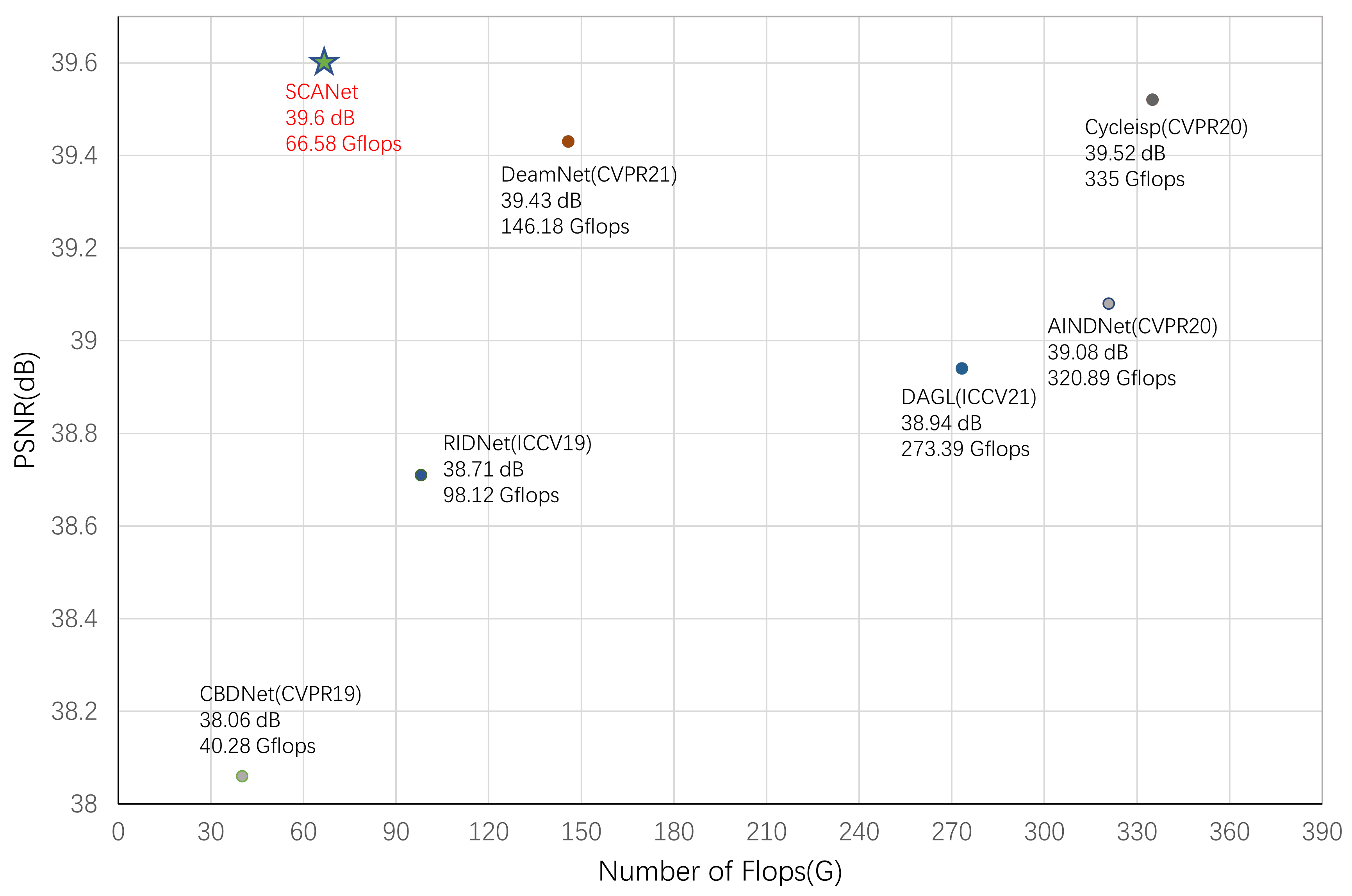}
    \caption{PSNR at different computational cost  amount of our method and previous excellent performance methods in SIDD dataset. Flops is computed on an image size of 256×256. }
    \label{fig-flops-compare}
\end{figure}

We demonstrate the effectiveness and efficiency of SCANet on various image denoising tasks. As shown in Fig.~\ref{fig-flops-compare}, SCANet outperforms the previous state-of-the-art (Cycleisp\cite{zamir2020cycleisp}) by 0.08 dB and 0.09 dB on the SIDD\cite{abdelhamed2018high} and DND\cite{plotz2017benchmarking} benchmarks, respectively, while reducing the computational complexity by four-fifths. Finally, we summarize the contributions of this work as follows:

• We propose a efficient yet effective SCANet,  which is a novel  UNet structured network designed with dual branch for single image restoration.

• We propose a complementary attention mechanism CAM via dense-sparse module, which enables SCANet to achieve better image denoising performance.

• We design the gradient-based branch, which can effectively preserve the structural details via structure-preserving, and this branch can be embedded in any current image restoration network with simple modification.

• We extensively compare with different methods and show that our method outperforms the state-of-the-art on various open source real noise datasets.


\section{Related Works}

\subsubsection{Traditional Methods.}Image denoising is a basic task in image processing. Early works include  non-local means (NLM)\cite{buades2005non}, 3D transform domain filtering (BM3D)\cite{dabov2007image}, and others\cite{mairal2009non,portilla2003image,foi2007pointwise,xu2017multi,dong2012nonlocally}, these algorithms usually need to rely on image priors. Although classical methods can obtain reasonable denoising results, they usually have some problems, such as high computational complexity, poor generalization performance, poor blind denoising effect and so on. In recent years, with the development of CNN and deep learning, end-to-end training denoising CNN has made unprecedented progress in this field and is one of the hot research directions.

\subsubsection{Deep Learning.}Denoising based on deep learning is mainly to design  CNN architecture to solve. Earlier some work, trained a simple multilayer perceptron (MLP)\cite{burger2012image} on a large synthetic noise dataset, and this approach performed well compared to previous traditional algorithms. Recent denoising methods based on deep neural networks\cite{chen2016trainable,zamir2020learning,tai2017memnet,jain2008natural,zhou2020awgn,cheng2021nbnet,xie2012image,ulyanov2018deep,ren2018dn} usually learn image priors and noise distribution information in images through a large number of paired training datasets to obtain Better denoising,More and more advanced network architectures are emerging. DnCNN\cite{zhang2017beyond} demonstrated the effectiveness of residual learning and batch normalization using CNN. Later, more network structures were proposed to make improvements, such as dilated convolution\cite{wang2017dilated,wang2019multi}, autoencoder with skip connections\cite{mao2016image,liu2018denoising}, and attention mechanism\cite{woo2018cbam,hu2018squeeze,liu2020residual}. FFDNet\cite{zhang2018ffdnet}proposed to use noise estimation map as input to balance the suppression of uniform noise and the maintenance of details, so as to deal with more complex real scenes.Recently, there have been some methods of synthesizing real noisy data to enhance the performance of the model. Cycleisp\cite{zamir2020cycleisp} proposed to use neural network to simulate the forward and reverse process of Camera ISP to obtain more realistic synthetic noisy data and train the denoising network to obtain Advanced denoising performance. Deamnet\cite{ren2021adaptive} proposes to incorporate a new adaptive consistency prior term into the optimization problem, and then utilize the optimization process to inform deep network design by using an unrolling strategy, and enhances its interpretability to a certain extent. Peking University et al.\cite{mou2021dynamic} proposed an attention model DAGL based on dynamic graph neural network to study the dynamic non-local similarity at the block level in the process of image restoration.Furthermore, several recent methods using deep CNNs\cite{anwar2019real,chang2020spatial,yue2019variational,pang2021recorrupted,huang2021neighbor2neighbor,zheng2021deep} have also demonstrated promising denoising performance.


\section{Overall architecture for image denoising}

\DeclareGraphicsExtensions{.eps,.png,.jpg,.bmp}

\begin{figure}[h]
    \centering
    \includegraphics[width=12cm]{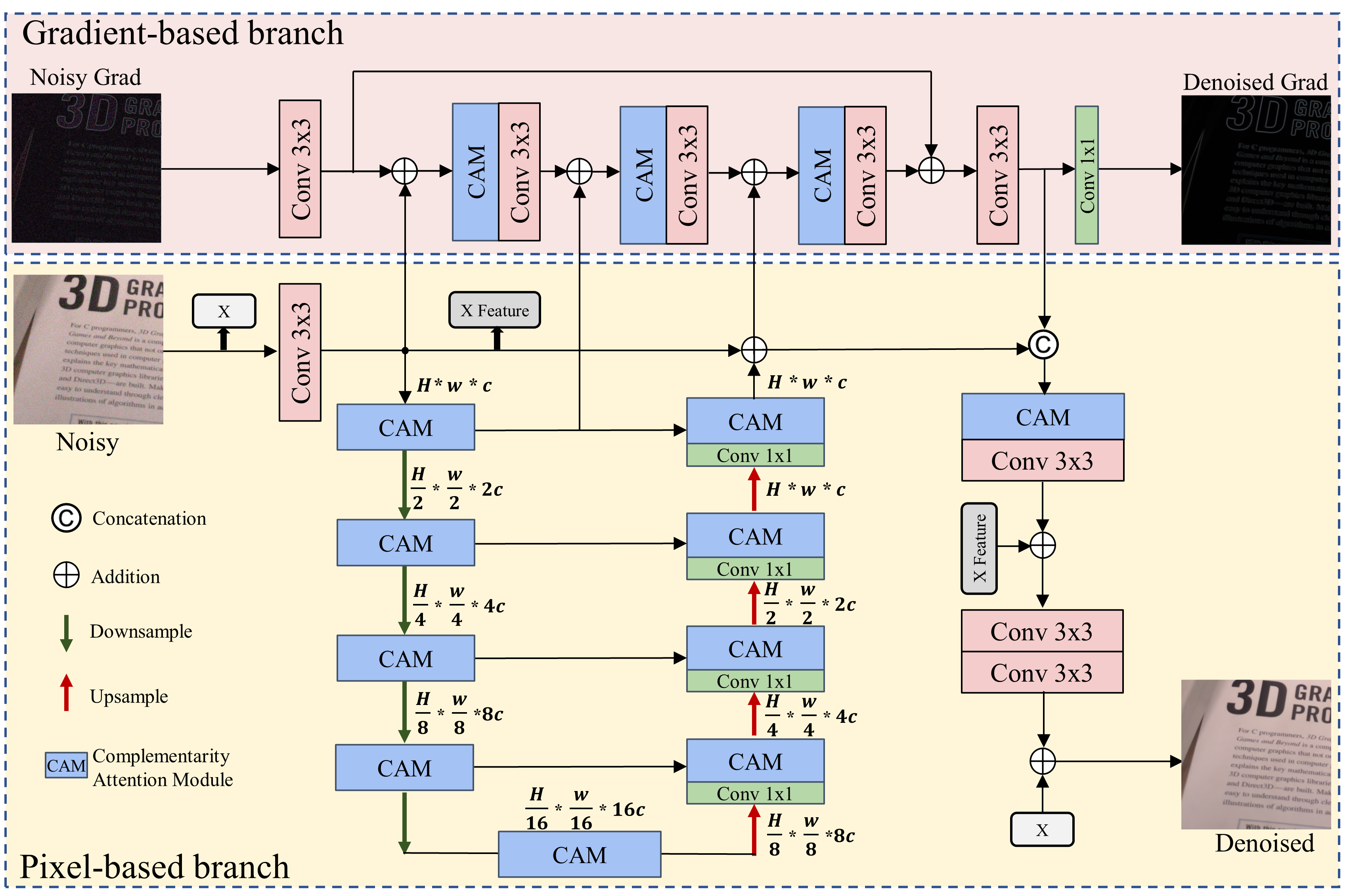}
    \caption{ Overall framework of our SCANet. Our architecture consists of two branches, the Pixel-based branch and the Gradient-based branch. The Gradient-based branch aims to input gradient maps to the ground-truth counterparts. }
    \label{fig-scaunet}
\end{figure}

Our main goal is to develop an efficient lightweight CNN model that can effectively deal with image denoising tasks. To effectively utilize dense and sparse features while keeping light weight to reduce computational complexity,and pay attention to the structural feature information of the image, we propose a Complementary Attention Module (CAM). We show the overall flow of our network architecture. The proposed denoising network is constructed from multiple CAMs, which mainly consists of two branches: (a) a pixel-based branch with a Unet structure. (b) a gradient-based branch with a cascaded residual structure. The overall framework is depicted in Fig.~\ref{fig-scaunet}.

\section{Proposed method}
\label{sec:blind}

\subsection{CAM: complementarity Attention Module}

\begin{figure}[H]
    \centering
    \includegraphics[width=12cm]{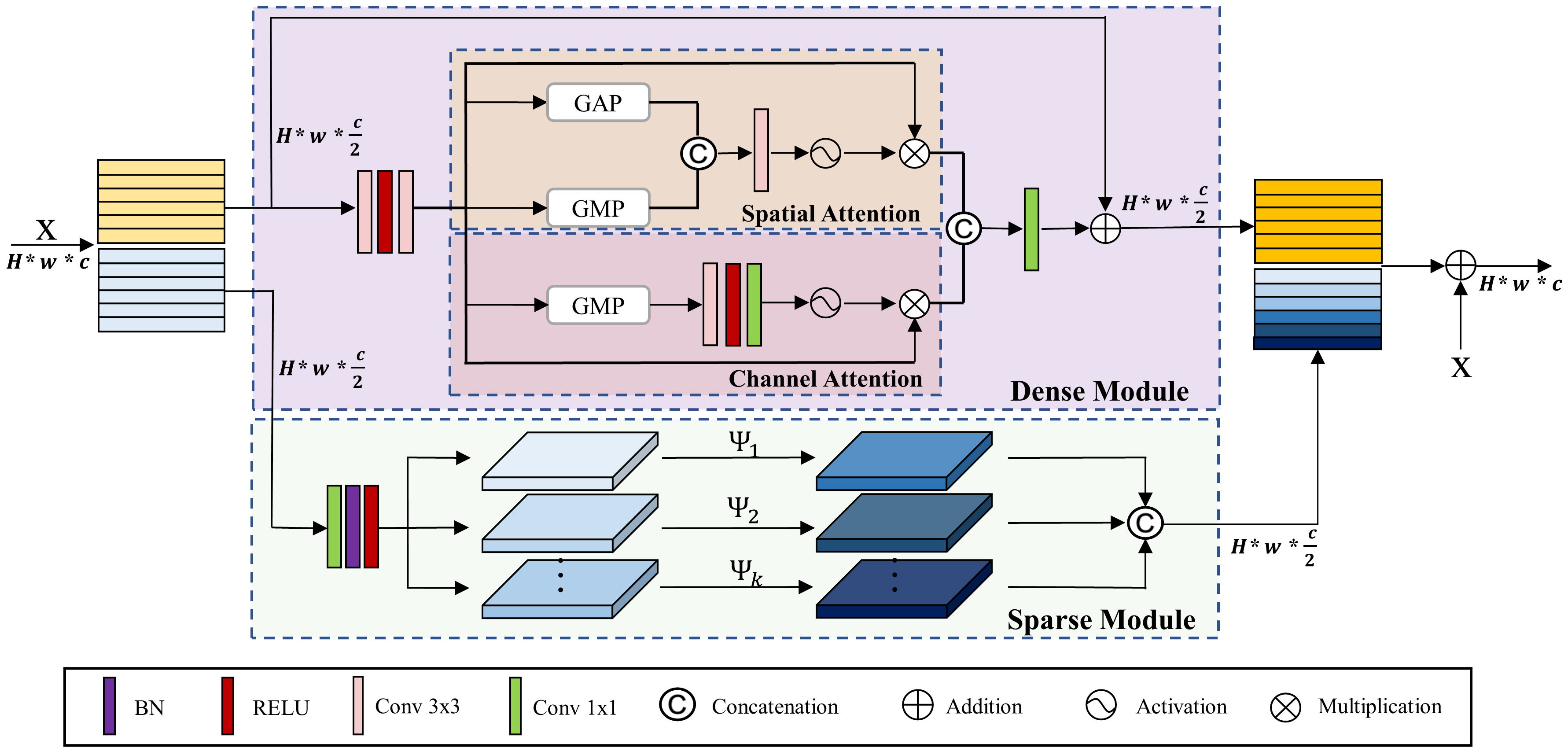}
    \caption{ Overall framework of our Complementarity Attention Module(CAM). \( \Psi \) represents the cheap operation. }
    \label{fig-cam}
\end{figure}

The proposed CAM aims to Mining the complementarity of dense and sparse features, suppress the useless feature in terms of both spatial-wise and channel-wise. As a result, more informative features can be propagated via a proposed module. Furthermore, CAM can be divided into Dense Module and Sparse Module. The overall framework of the proposed CAM architecture is shown in Fig.~\ref{fig-cam}.

\subsubsection{Dense Module(DM).}As shown in Fig.~\ref{fig-cam}, this module contains a dual attention branches: {\bfseries (a) Spatial Attention(SA).} This branch uses the correlation of feature mapping to calculate the spatial attention weight mapping, and then used to weight the incoming features $U$. We first perform global average pooling and max pooling operations on the features separately in the channel dimension, and concatenate the output feature maps. Next is convolution and sigmoid activation to obtain spatial attention map.  {\bfseries (b) Channel Attention(CA).} This branch aims to capture the channels correlation of convolution features. We first perform a compression operation to encode the global feature map, and then perform a sigmoid activation to fully capture the relationship between channels. The overall dense module can be defined as:
\begin{align}
Y_{D M}=X_{i n}+M_{c}([{F_{SA}}(U), {F_{CA}}(U)])
\end{align}
where $U$ denotes features maps that are obtained by applying convolution block on input features $X_{in}$, and $M_{c}$ is the last $1\times1$ Convolution of fused features, $F_{SA}$ and $F_{CA}$ denote the spatial attention mechanism and the channel attention mechanism, respectively.

\subsubsection{Sparse Module(SM).}Deep CNN is usually composed of a large number of convolutions, with a huge amount of computation, but there is a lot of redundancy in its intermediate feature mapping. By establishing the attention mechanism of group convolution, we use some different cheap linear operators to replace the ordinary convolution. Then, these features are executed on each channel to generate sparse features with low computational complexity. Compared with ordinary convolution, this module can significantly reduce the amount of computation. Specifically, it is assumed that the output feature mapping is the mapping of several other inherent feature mappings, with some cheap transformations, that is, redundant feature mapping can be transformed from other feature mapping.

As shown in Fig.~\ref{fig-cam}, we first transform the input features through a convolution block to intermediate features, it is smaller, and the calculation cost is also less.Then we perform a series of cheap linear operations on each intermediate features to obtain the desired n feature maps, combine the two to get the output feature map. Note that the linear operation \( \Psi \) calculation quantity less on each channel than conventional convolution. The overall process is:
\begin{align}
Y= f_{conv}(X)
\end{align}
\begin{align}
y_{i j}=\Psi_{ij}\left(y_{i}\right), \quad \forall i=1, \ldots, c, \quad j=1, \ldots, s
\end{align}

where $f_{conv}$ is the convolution block, $Y$ $\in$ $\mathbb{R}^{h \times w \times c}$ is the intermediate feature map with n channels, \( y_{i} \) is the i-th intermediate feature map in $Y$, \( \Psi_{i j} \)  in the above function is the j-th linear operation for generating the j-th sparse feature map \( y_{i j} \). So, \( y_{i} \) can have one or more Similar sparse feature maps $\left\{y_{i j}\right\}_{j=1}^{s}$. By utilizing Eq. 3, we can obtain $c_{out}$ = c $*$ s $+$ c sparse feature maps $\left\{y_{i j}\right\}_{i=1,j=1}^{c,s}$ as the output of a sparse module as shown in Fig.~\ref{fig-cam}.

\subsection{Gradient-based Branch}
The purpose of the gradient-based branch is to transform the input noise gradient map into the corresponding denoised gradient map, and learn the mapping between the two modes by using image to image conversion technology. Using the characteristics of gradient map, the neural network can pay more attention to the spatial relationship of high-frequency details and capture the structural correlation, so as to generate the approximate gradient map of the denoised image, get better denoising effect via structure-preserving. The method of obtaining the gradient map of image I is:
\begin{align}
\begin{aligned}
I_{i}(x) &=I(i+1, j)-I(i-1, j) \\
I_{j}(x) &=I(i, j+1)-I(i, j-1) \\
grad I(x) &=\left(I_{i}(x), I_{j}(x)\right) \\
G \odot I &=\|grad I({x})\|_{2}
\end{aligned}
\end{align}

where $G$ stands for the operation to extract gradient map whose elements are gradient lengths for pixels with coordinates ${x} = (i, j)$. As shown in Fig.~\ref{fig-scaunet}, since the rich structural information contained in the pixel-based branch is important for the restoration of the gradient map, the gradient-based branch introduces the output features of several intermediate layers from the pixel-based branch, and we fuse these features to improve the performance of the gradient-based branch, and can effectively reduce the redundant design of the model. We use CAM as basic block to build gradient-based branch. Through the gradient feature mapping information obtained from the gradient-based branch, we finally integrate them into the pixel-based branch, so as to  feed back and guide the image denoising as a structural priori. Finally, we generate the denoise gradient maps by a $1\times1$ convolution layer.

\subsection{Loss Functions}

\subsubsection{Pixel Loss.}
The network is optimized by a pixel loss. This metric loss can reduce the average pixel difference between the restored image and the real image and is widely used to improve model fitting and denoising performance. We use improved charbonnierloss\cite{lai2018fast} based on l1 as the pixel loss function, written as:
\begin{align}
\mathcal{L}_{Pixel}=\sqrt{\left\|I_{Denoised}-I_{GT}\right\|+\epsilon}
\end{align}

\subsubsection{Gradient Loss.}
 We use the corresponding gradient loss to achieve the learning of gradient information. Our gradient loss optimization goal is to reduce the gap between the output denoised gradient map and the gradient maps extracted from the corresponding ground-truth images. Under the supervision of both the image and gradient space, the model takes care to avoid the loss of high-frequency details while ensuring denoising. Therefore, we design two loss terms to penalize the difference in the gradient map of denoised and groundtruth images, as follows:
\begin{align}
\mathcal{L}_{Pixel-based}^{Grad}=\mathbb{E}_{{L1}}\left\|G\odot I_{Denoised}-G\odot I_{GT}\right\|
\end{align}

The other one loss term to penalize the difference between the gradient-based branch output and the gradient map (GM) of the real image, to reconstruct high-quality gradient maps by minimizing the pixelwise loss as follows:
\begin{align}
\mathcal{L}_{Gradient-based}^{Grad}=\mathbb{E}_{{L2}}\left\|I_{Gradient}^{GM}-G\odot I_{GT}\right\|
\end{align}

\subsubsection{Overall Objective Loss.}
Gradient-based branch incorporates multi-level representations from the pixel-based branch to improve performance, reduce parameters and outputs gradient information to guide the denoise process by a fusion block in turn. The final denoise outputs are optimized by not only conventional image-space losses, but also the proposed gradient-space objectives. In conclusion, we have two branch tasks, but the optimizations of the two branches are related to each other.The loss function for the overall objective is :

\begin{equation}
\begin{aligned}
\mathcal{L}_{Denoise }&=\mathcal{L}_{Pixel-based }+\mathcal{L}_{Gradient-based } \\
&=\alpha \mathcal{L}_{Pixel }+\beta \mathcal{L}_{Pixel-based }^{Grad }+\gamma \mathcal{L}_{Gradient-based }^{Grad }
\end{aligned}
\end{equation}

 $\alpha$, $\beta$ and $\gamma$ denote the weight parameters of different losses, $\alpha$ is the weights of the pixel losses, $\beta$ and $\gamma$ are the weights of the pixel-based and gradient-based losses, respectively.


\section{Experiments}
\subsection{Real Image Datasets}

\subsubsection{SIDD dataset\cite{abdelhamed2018high}.}The dataset is based on 5 cameras (Google Pixel, iPhone 7, Samsung Galaxy S6 Edge, Motorola Nexus 6, LG G4) shooting 10 scenes under four camera parameters, 200 scene instances, each scene is shot continuously 150 images. Among them, 160 scene instances are used as the training set, and 40 scene instances are used as the test set. There are 320 image pairs available for training and 1280 image pairs for validation.

\subsubsection{DND dataset\cite{plotz2017benchmarking}.}The dataset consists of 50 pairs of noisy and noise-free images captured by four consumer cameras. Shoot 50 scenes, including indoor and outdoor scenes. The provider extracts 20 crops of size $512\times512$ from each image, resulting in a total of 1000 patches. The full dataset is used for testing, and the real noise-free images are not publicly available. Quantitative evaluation in terms of PSNR and SSIM is only possible through the online server.

\subsection{Implementation Details}
We choose PSNR and SSIM as evaluation metrics, with higher values indicating better denoising performance. The models presented in this paper are trained with Adam\cite{da2014method} optimizer ($\beta1 = 0.9, \beta2 = 0.999$) and image crops of $128 \times 128 $. We randomly perform horizontal, vertical flips and rotation as data augmentation, we train for 60 epochs, batch size of 16, and initial learning rate of $1e-4$ which is decreased by a factor of 10 after every 25 epoch.As for the weight parameter of the loss, set $\alpha$, $\beta$ and $\gamma$ to 1, 0.1 and 0.2 for better performance of gradient translation. All the experiments are implemented by PyTorch\cite{paszke2019pytorch} on NVIDIA M40 GPUs, FLOPs and runtime is computed on an image size of 256×256.

\subsection{Ablation Study}
We analyze each component of SCANet for image denoising on the SIDD dataset. Table 1 shows the ablation results, it can be seen that our proposed CAM  and gradient-base branch can effectively improve the denoising performance. We can see from the ablation experiment of CAM that both the Dense Module and the Sparse Module we proposed can effectively improve the denoising performance of the network, and the Sparse Module can improve the performance with the lower computational cost. The complete application of CAM can maximize the denoising performance of the model.

\begin{figure}[t]
    \centering
    \includegraphics[width=12cm]{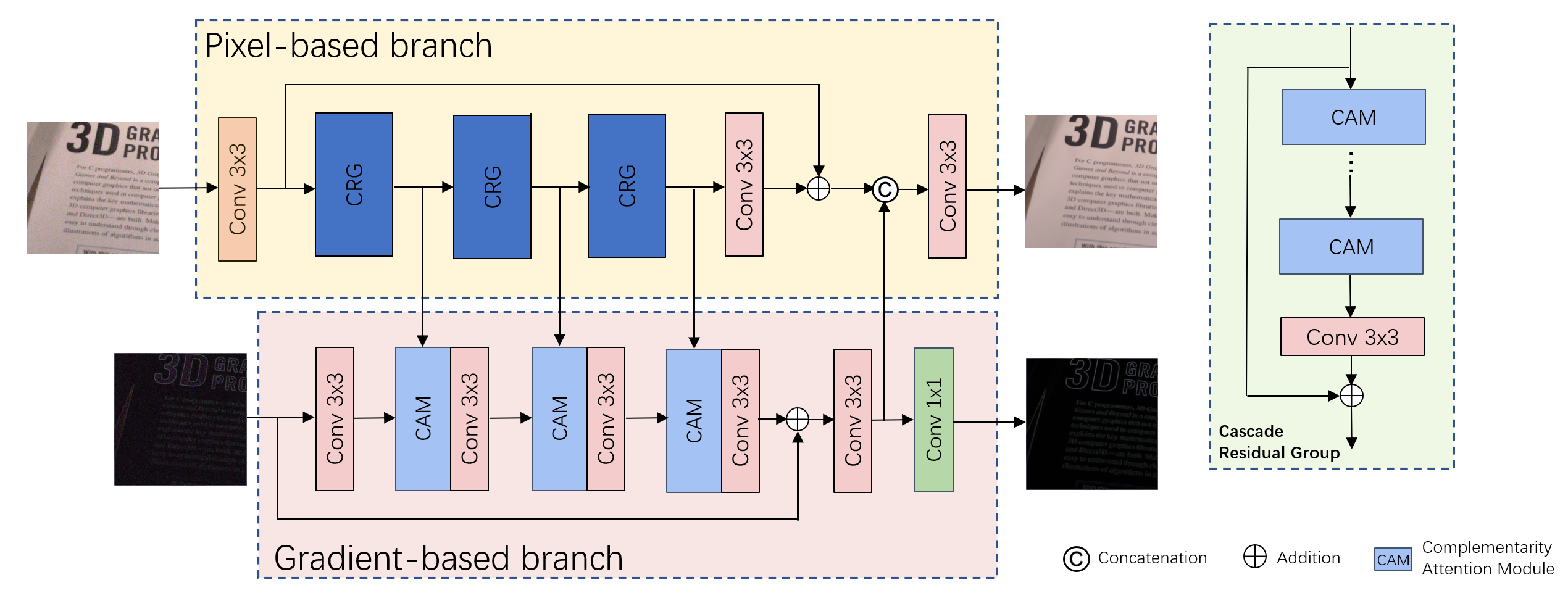}
    \caption{ Overall framework of Cascade structure model for denoising.}
    \label{fig-cascade}
\end{figure}

In addition, we also built a cascaded structured network based on CAM to test the performance impact of the proposed components under different structured networks, as shown in Fig.~\ref{fig-cascade}. Experiments show that the performance can be improved by 0.4dB and 0.45dB by integrating different components in cascade structure and UNet structure, respectively. And through the comparison, it can be seen that the proposed CAM  and gradient-based branch can achieve better denoising performance and lower inference time than the conventional cascade structure in a Unet structure of multiple scales.

\setlength{\tabcolsep}{4pt}
\begin{table}[H]
\begin{center}
\caption{Ablation study on other modules}
\label{table:Ablation study on other modules}
\scriptsize  
\begin{tabular}{lcccccc}
\hline\noalign{\smallskip}
Model & DM & SM & Grad-branch & Flops(G) & Runtime(ms) & PSNR(dB) \\
\noalign{\smallskip}
\hline
\noalign{\smallskip}
Cascade structure   & & & & 15.92  &  11.3 & 38.75  \\
Cascade structure  & \checkmark & & &  18.96 & 24.1   & 38.88  \\
Cascade structure  &\checkmark & \checkmark & &  27.26  &  44.3 & 39.06  \\
Cascade structure  & \checkmark & \checkmark & \checkmark &  49.73  &  72.6  & 39.15  \\
\noalign{\smallskip}
\hline
\noalign{\smallskip}
SCANet  & & & &  17.99 & 11.2  & 39.04  \\
SCANet  & \checkmark & & &  21.37 & 18.2  & 39.24  \\
SCANet  &\checkmark & \checkmark & &  45.65 &  34.3   & 39.52  \\
SCANet  & \checkmark & \checkmark & \checkmark &  66.58  &  \bf{54.7} & \bf{39.60}  \\

\hline
\end{tabular}
\end{center}
\end{table}
\setlength{\tabcolsep}{4pt}

\subsubsection{Effectiveness  of the Gradient-based Branch.}
\begin{figure}[h]
    \centering
    \includegraphics[width=12cm]{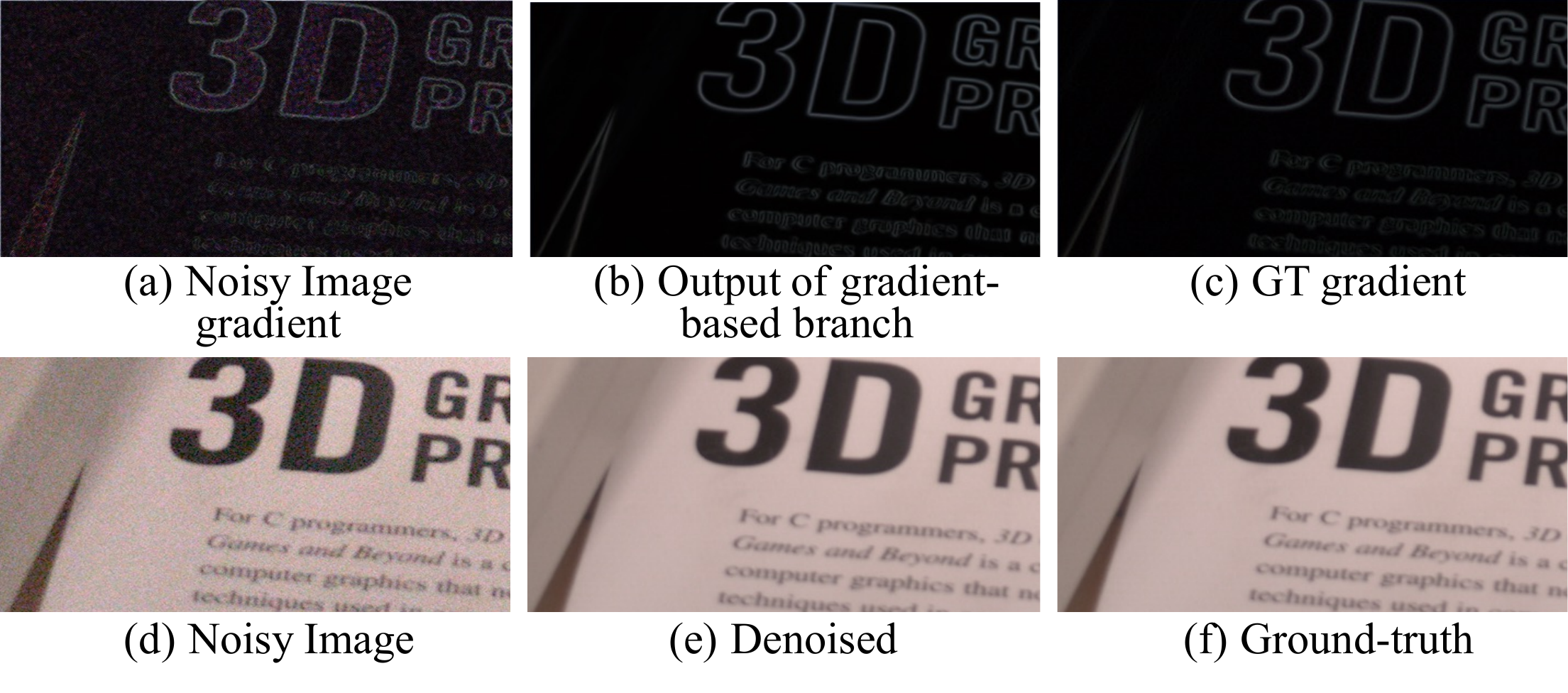}
    \caption{Visualization of the gradient map (image from sidd). Our gradient-based branch is able to recover denoised gradient maps with more structural details. }
    \label{fig-grad_vis}
\end{figure}

\begin{figure}[H]
\makeatletter
\makeatother
\centering
\subfigure [Only the pixel-based branch] {\includegraphics[width=6cm]{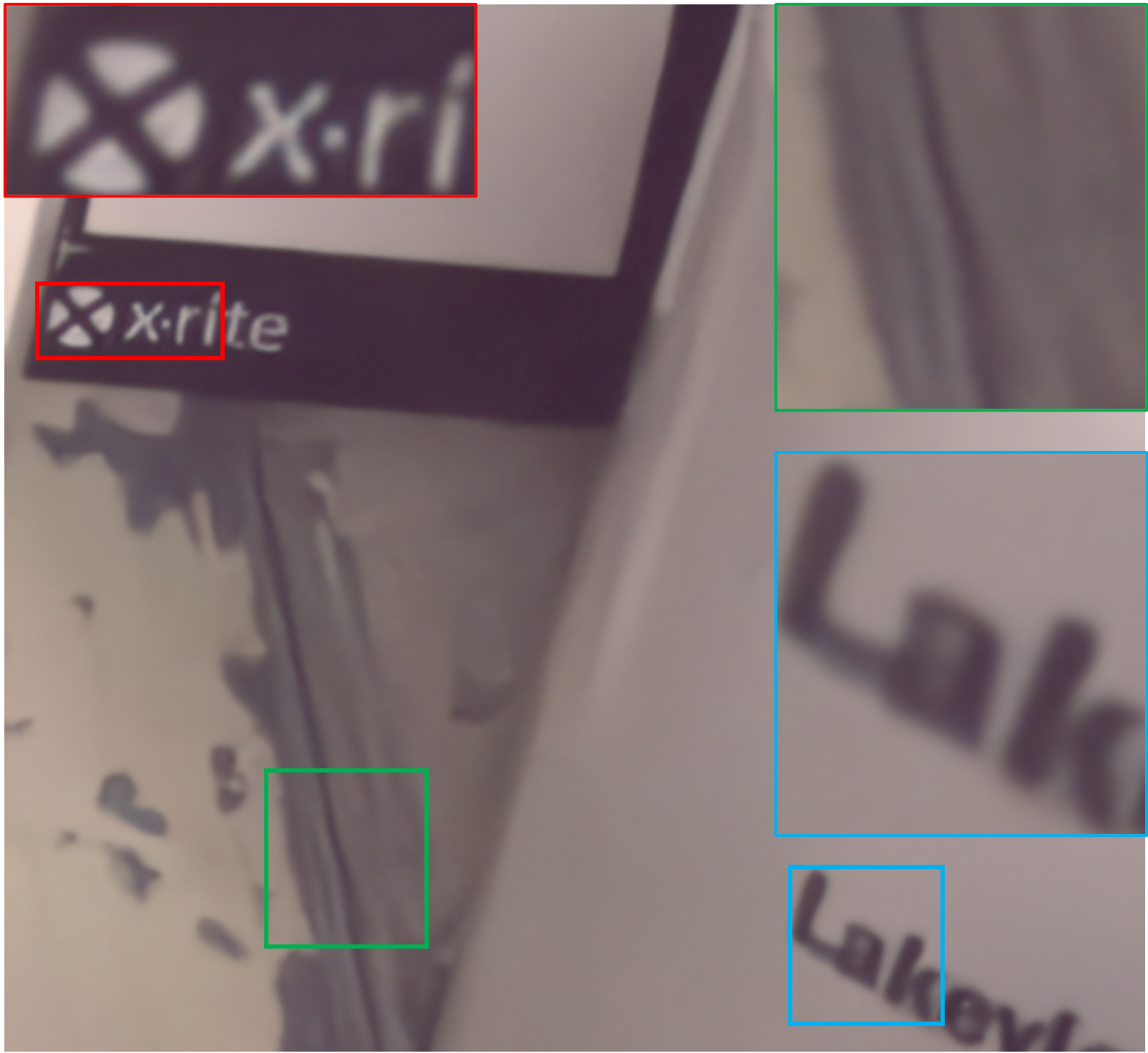}}
\centering
\subfigure [Complete model] {\includegraphics[width=6cm]{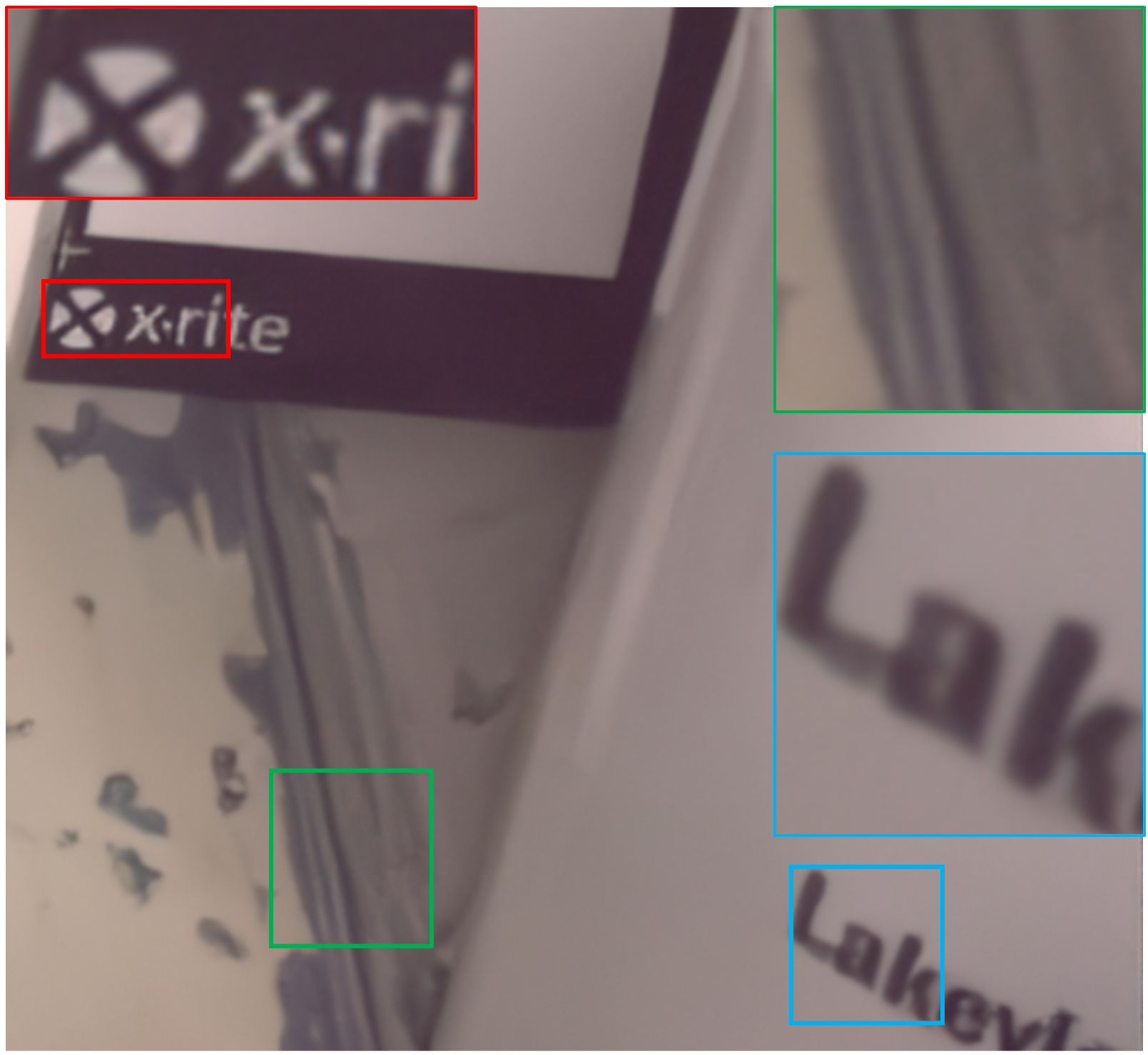}}
\caption{Denoise comparison of the models without and with the gradient-based branch.(image from sidd dataset). The image restored by the complete model has clearer structural texture details.} 
\label{fig-compare-grad}
\end{figure}

We visualize the output gradient graph to verify the effectiveness of gradient-based branch, as shown in Figure 6. For the input image with serious noise, the extracted gradient map will also contain serious noise interference; in the same way, the corresponding gradient map extracted from the GT image with clear contour details also has detailed and clear contours.
From the output gradient map in Fig.~\ref{fig-grad_vis}(b), it can be seen that the gradient-based branch successfully recovers denoised and well-structured gradient maps, and can provide clear structural information guidance for pixel-based branch. We conduct experiments to evaluate the effectiveness of the gradient-based branch, we only use the pixel-based branch for inference. The visualization results are shown in Fig.~\ref{fig-compare-grad}, we can see that the pictures and texts denoised only by the pixel-based branch are more blurry than those recovered by the complete model. Differences in detail textures suggest that gradient-based branch can help preserve sharp structural details for better denoising.

\subsection{Results for Denosing}
\begin{figure}[H]
\makeatletter
\makeatother
\centering
\subfigure  {\includegraphics[width=12cm]{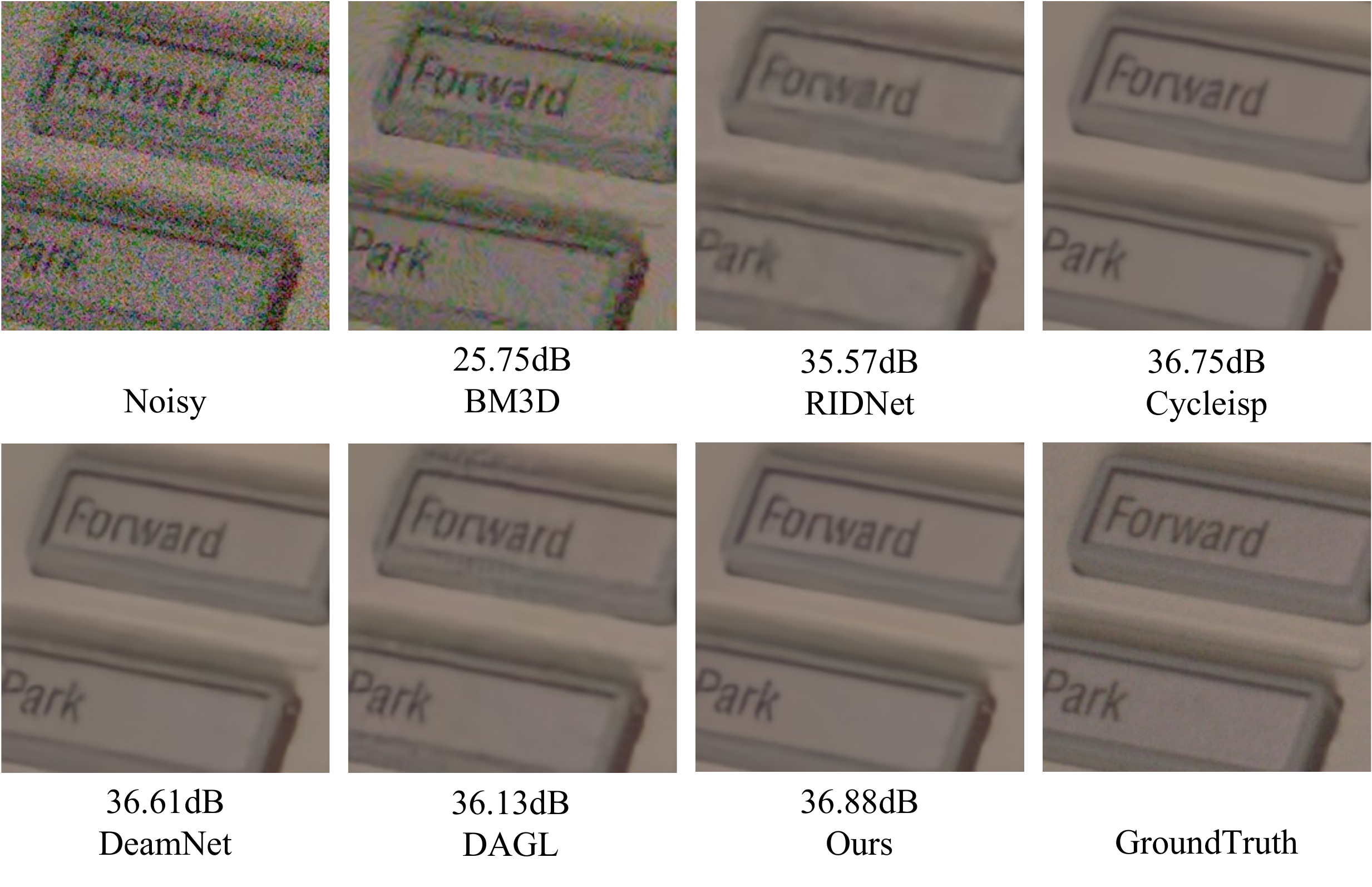}}
\centering
\subfigure {\includegraphics[width=12cm]{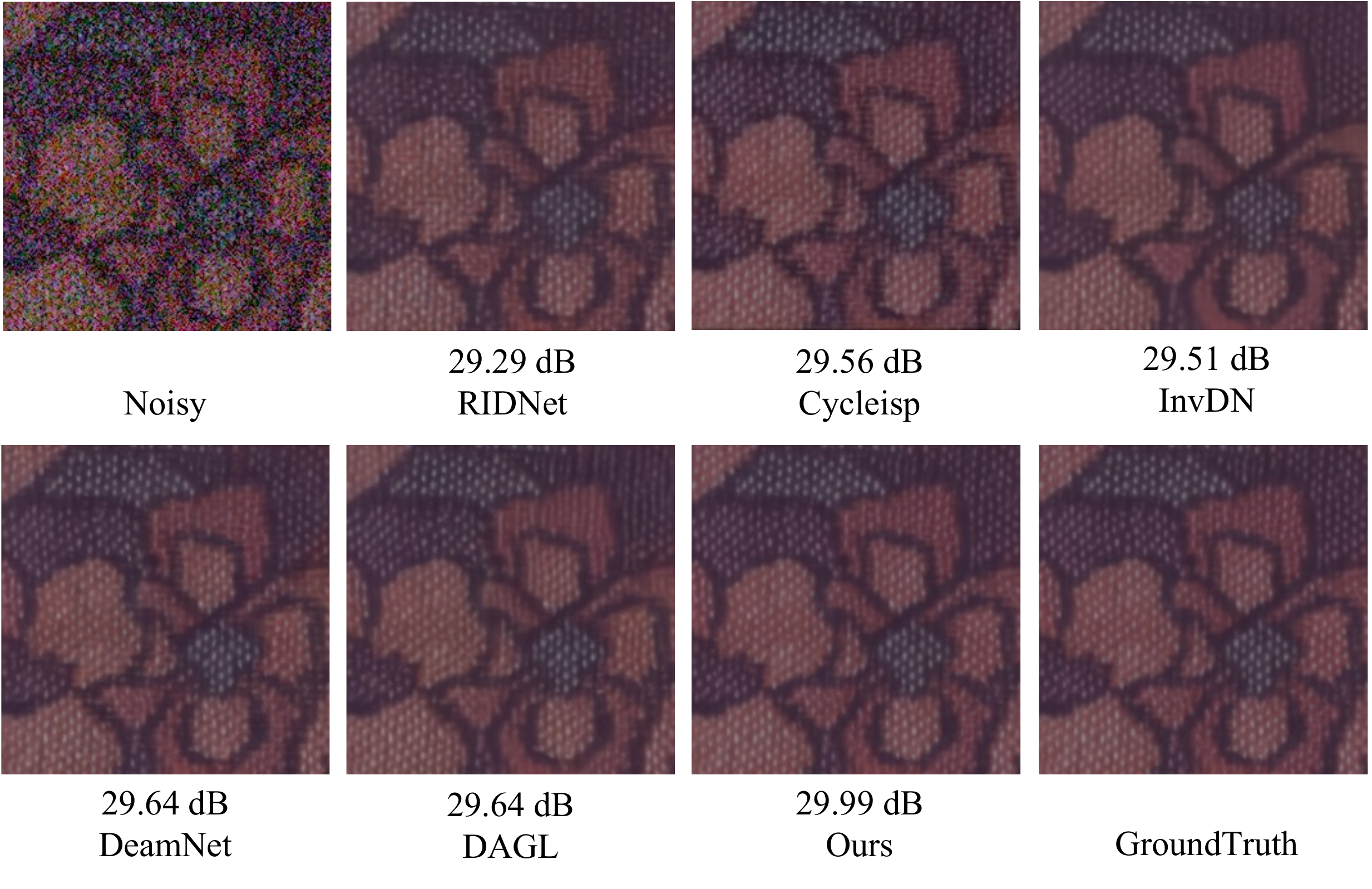}}
\caption{Denoising results of different methods on  sRGB images from the SIDD dataset.}
\label{Denosing_for_sidd}
\end{figure}

\begin{figure}[H]
\makeatletter
\makeatother
\centering
\subfigure  {\includegraphics[width=12cm]{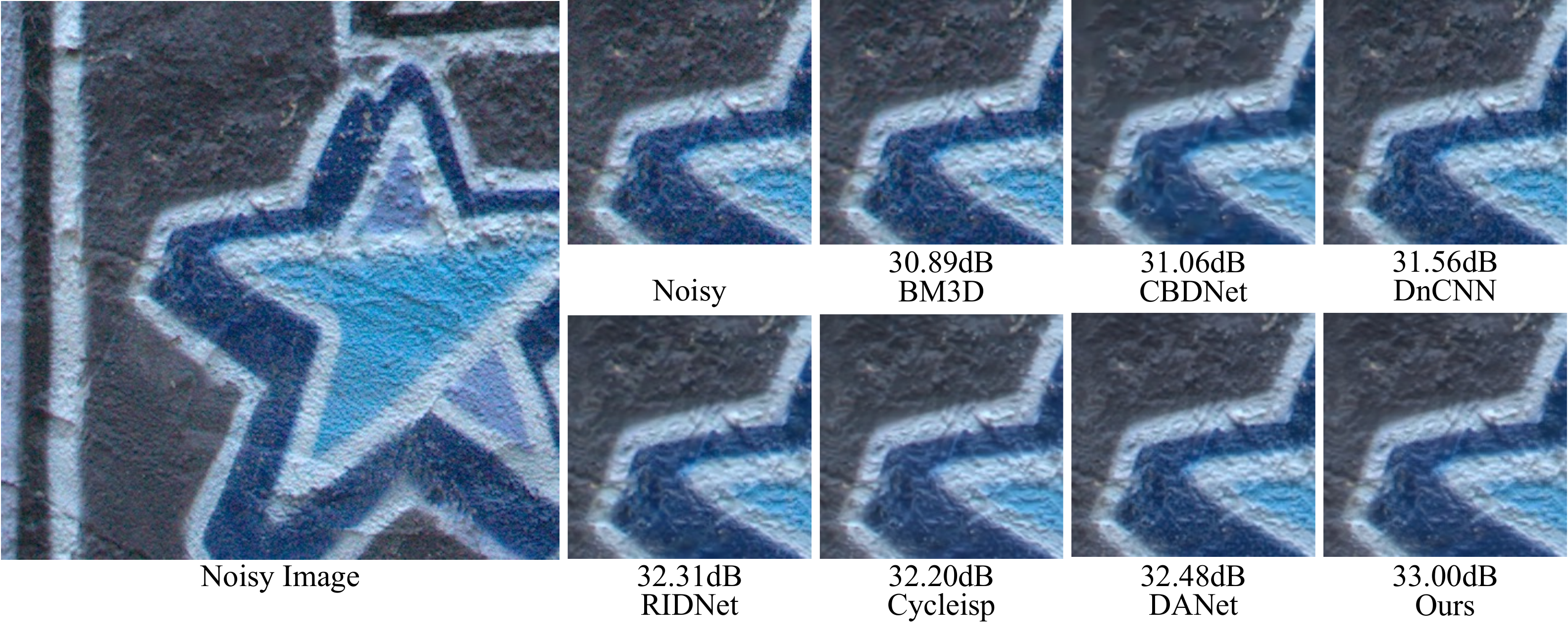}}
\centering
\subfigure {\includegraphics[width=12cm]{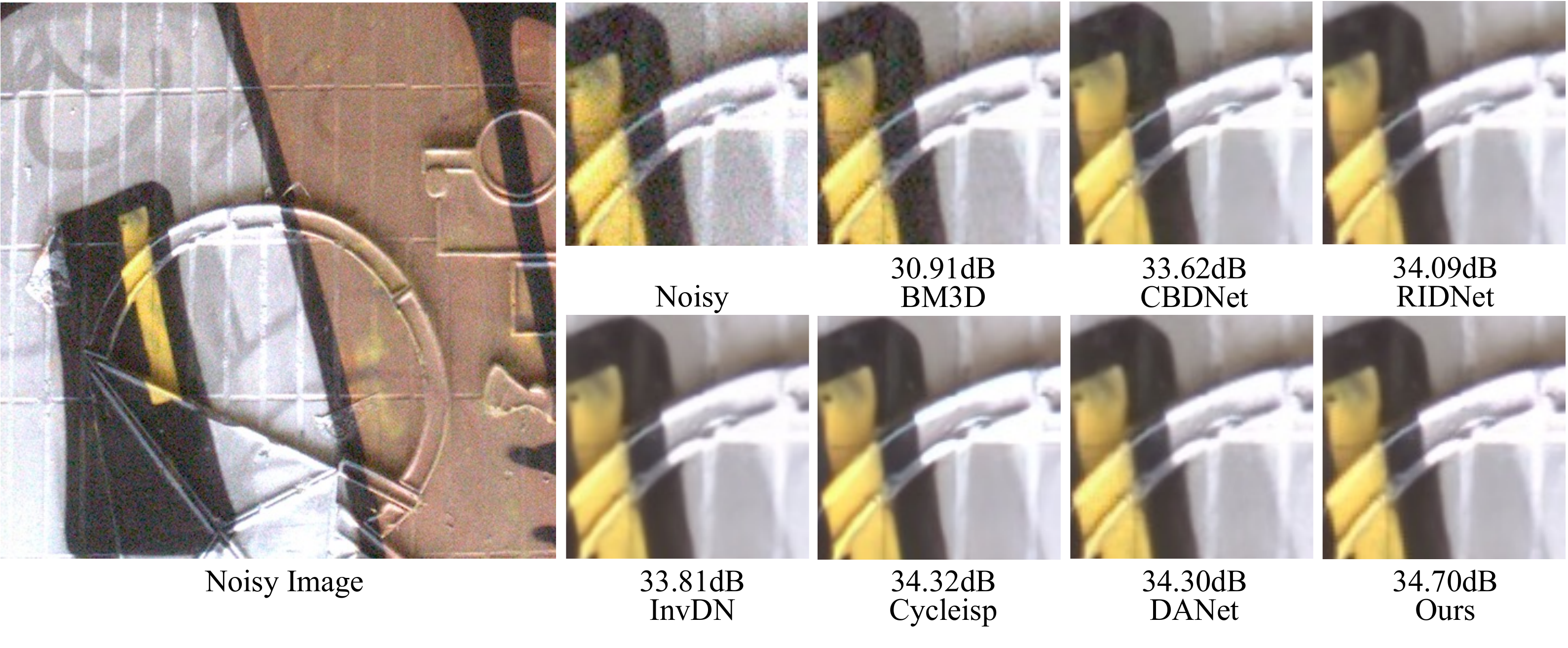}}
\caption{Denoise sRGB images from DND. Our method recovers fine edge details very clearly.}
\label{Denosing_for_dnd}
\end{figure}

We compare the denoising results of different methods on sRGB images from the DND and SIDD datasets, including the traditional method and the outperforming CNN algorithm in recent papers. Table ~\ref{table:Results for Denosing in dnd} and Table ~\ref{table:Results for Denosing in sidd} show the scores of image quality metrics. Overall, the proposed model performs favorably against the state-of the-art, and the model requires less computation. And we compared the computational complexity and runtime of the model under different denoising methods, as shown in Table ~\ref{table:Results for Analysis of the complexity}. It can be seen that our method achieves the best computational efficiency. Compared to the Cycleisp\cite{zamir2020cycleisp}, our approach demonstrates the performance gain of 0.08 dB and 0.09 dB on SIDD and DND datasets respectively, and our model is only one-fifth of its computational cost.
Fig.~\ref{Denosing_for_sidd} and Fig.~\ref{Denosing_for_dnd} illustrate the sRGB denoising results on SIDD and DND, respectively. To denoise, most of the evaluated algorithms either produce over-smooth images (and sacrifice image details) or generate images with splotchy texture and chroma artifacts. In contrast, our method generates clean and artifact-free results, while the image details are effectively preserved.

\setlength{\tabcolsep}{1.4pt}
\begin{table}[htbp]
\begin{center}
\caption{Denoising sRGB images of the DND dataset}
\label{table:Results for Denosing in dnd}
\scriptsize  
\begin{tabular}{lcccccccccccc}
\hline\noalign{\smallskip}
Method   & BM3D & KSVD & WNNM  & CBDNet & RIDNet & VDNet &AINDNet & Cycleisp & DeamNet & InvDN & SCANet\\
        & \cite{dabov2007image} & \cite{aharon2006k} & \cite{gu2014weighted} &   \cite{guo2019toward} & \cite{anwar2019real} & \cite{yue2019variational} & \cite{kim2020transfer}&  \cite{zamir2020cycleisp}   & \cite{ren2021adaptive} & \cite{liu2021invertible}   \\
\noalign{\smallskip}
\hline
\noalign{\smallskip}

PSNR    & 34.51 & 36.49 & 34.67  & 38.06 & 39.23 & 39.38 & 39.53 & 39.56  & 39.63 & 39.57& \bf{39.65}\\
SSIM    & 0.851 & 0.898 & 0.865  & 0.942 & 0.953 & 0.952 & 0.956 & 0.956  & 0.953 & 0.952 & \bf{0.953} \\
\hline
\end{tabular}
\end{center}
\end{table}
\setlength{\tabcolsep}{1.4pt}

\setlength{\tabcolsep}{1.4pt}
\begin{table}[htbp]
\begin{center}
\caption{Denoising sRGB images of the SIDD dataset}
\label{table:Results for Denosing in sidd}
\scriptsize  
\begin{tabular}{lccccccccccc}
\hline\noalign{\smallskip}
Method & DnCNN  & BM3D  & NLM & CBDNet & RIDNet &AINDNet& Cycleisp & DAGL & InvDN & DeamNet & SCANet\\
       & \cite{zhang2017beyond}  & \cite{dabov2007image}  & \cite{buades2005non} & \cite{guo2019toward} & \cite{anwar2019real} &   \cite{kim2020transfer}& \cite{zamir2020cycleisp} & \cite{mou2021dynamic} & \cite{liu2021invertible} & \cite{ren2021adaptive}    \\
\noalign{\smallskip}
\hline
\noalign{\smallskip}

PSNR  & 23.66  & 25.65  & 26.76 & 30.78 & 38.71 & 39.08 & 39.52 & 38.94 & 39.28 & 39.35 & \bf{39.60}\\
SSIM  & 0.583  & 0.685  & 0.699 & 0.754 & 0.914 & 0.953& 0.957 & 0.953 & 0.955 & 0.955 & \bf{0.969} \\
\hline
\end{tabular}
\end{center}
\end{table}
\setlength{\tabcolsep}{1.4pt}

\setlength{\tabcolsep}{1.4pt}
\begin{table}[htbp]
\begin{center}
\caption{Analysis of the complexity and inference speed of different models}
\label{table:Results for Analysis of the complexity}
\scriptsize  
\begin{tabular}{lcccccccccc}
\hline\noalign{\smallskip}
Method      & RIDNet & AINDNet & Cycleisp & DAGL& InvDN & DeamNet & SCANet\\
         & \cite{anwar2019real} & \cite{kim2020transfer} &  \cite{zamir2020cycleisp}  & \cite{mou2021dynamic} &\cite{liu2021invertible}& \cite{ren2021adaptive}    \\
\noalign{\smallskip}
\hline
\noalign{\smallskip}

FLOPS(G)    & 98.12 & 320.89  & 335.01 & 273.39 & 23.77 & 146.18 & \bf{66.58} \\
RUNTIME(ms)      &62.02& 82.67& 121.3 & 2856 & 62.79 & 87.92  & \bf{54.7} \\
\hline
\end{tabular}
\end{center}
\end{table}
\setlength{\tabcolsep}{1.4pt}

\section{Conclusions}

In this paper, we propose an efficient dual branch Unet structured image denoising framework with dense-sparse complementary attention via Structure-preserving. We introduce key designs into the core components of the cnn network to improve the aggregation transformation of dense and sparse features. Specifically, our proposed CAM contains two complementary attention mechanisms. The dense module can effectively learn feature information in both spatial and channel dimensions; the sparse module increases the feature expression ability of the model and replaces the conventional convolution, thus has linear complexity instead of quadratic complexity. Furthermore, we introduce a structural priori to the denoising process by constructing gradient-based branch, which effectively preserves clear image details. Finally, we conduct qualitative experiments on two publicly available real noise benchmark test sets, and the results demonstrate the effectiveness of our proposed method. Benefiting from efficient model structure, CAM and structure-preserving, Our method achieves better denoising performance at less computational cost.

\bibliographystyle{splncs}
\bibliography{egbib}

\begin{thebibliography}{10}

\bibitem{ronneberger2015u}
Ronneberger, O., Fischer, P., Brox, T.:
\newblock U-net: Convolutional networks for biomedical image segmentation.
\newblock In: International Conference on Medical image computing and
  computer-assisted intervention, Springer (2015)  234--241

\bibitem{zamir2020cycleisp}
Zamir, S.W., Arora, A., Khan, S., Hayat, M., Khan, F.S., Yang, M.H., Shao, L.:
\newblock Cycleisp: Real image restoration via improved data synthesis.
\newblock In: Proceedings of the IEEE/CVF Conference on Computer Vision and
  Pattern Recognition. (2020)  2696--2705

\bibitem{abdelhamed2018high}
Abdelhamed, A., Lin, S., Brown, M.S.:
\newblock A high-quality denoising dataset for smartphone cameras.
\newblock In: Proceedings of the IEEE Conference on Computer Vision and Pattern
  Recognition. (2018)  1692--1700

\bibitem{plotz2017benchmarking}
Plotz, T., Roth, S.:
\newblock Benchmarking denoising algorithms with real photographs.
\newblock In: Proceedings of the IEEE conference on computer vision and pattern
  recognition. (2017)  1586--1595

\bibitem{buades2005non}
Buades, A., Coll, B., Morel, J.M.:
\newblock A non-local algorithm for image denoising.
\newblock In: 2005 IEEE Computer Society Conference on Computer Vision and
  Pattern Recognition (CVPR'05). Volume~2., IEEE (2005)  60--65

\bibitem{dabov2007image}
Dabov, K., Foi, A., Katkovnik, V., Egiazarian, K.:
\newblock Image denoising by sparse 3-d transform-domain collaborative
  filtering.
\newblock IEEE Transactions on image processing \textbf{16} (2007)  2080--2095

\bibitem{mairal2009non}
Mairal, J., Bach, F., Ponce, J., Sapiro, G., Zisserman, A.:
\newblock Non-local sparse models for image restoration.
\newblock In: 2009 IEEE 12th international conference on computer vision, IEEE
  (2009)  2272--2279

\bibitem{portilla2003image}
Portilla, J., Strela, V., Wainwright, M.J., Simoncelli, E.P.:
\newblock Image denoising using scale mixtures of gaussians in the wavelet
  domain.
\newblock IEEE Transactions on Image processing \textbf{12} (2003)  1338--1351

\bibitem{foi2007pointwise}
Foi, A., Katkovnik, V., Egiazarian, K.:
\newblock Pointwise shape-adaptive dct for high-quality denoising and
  deblocking of grayscale and color images.
\newblock IEEE transactions on image processing \textbf{16} (2007)  1395--1411

\bibitem{xu2017multi}
Xu, J., Zhang, L., Zhang, D., Feng, X.:
\newblock Multi-channel weighted nuclear norm minimization for real color image
  denoising.
\newblock In: Proceedings of the IEEE international conference on computer
  vision. (2017)  1096--1104

\bibitem{dong2012nonlocally}
Dong, W., Zhang, L., Shi, G., Li, X.:
\newblock Nonlocally centralized sparse representation for image restoration.
\newblock IEEE transactions on Image Processing \textbf{22} (2012)  1620--1630

\bibitem{burger2012image}
Burger, H.C., Schuler, C.J., Harmeling, S.:
\newblock Image denoising: Can plain neural networks compete withfbm3 bm3d?
\newblock In: 2012 IEEE conference on computer vision and pattern recognition,
  IEEE (2012)  2392--2399

\bibitem{chen2016trainable}
Chen, Y., Pock, T.:
\newblock Trainable nonlinear reaction diffusion: A flexible framework for fast
  and effective image restoration.
\newblock IEEE transactions on pattern analysis and machine intelligence
  \textbf{39} (2016)  1256--1272

\bibitem{zamir2020learning}
Zamir, S.W., Arora, A., Khan, S., Hayat, M., Khan, F.S., Yang, M.H., Shao, L.:
\newblock Learning enriched features for real image restoration and
  enhancement.
\newblock In: European Conference on Computer Vision, Springer (2020)  492--511

\bibitem{tai2017memnet}
Tai, Y., Yang, J., Liu, X., Xu, C.:
\newblock Memnet: A persistent memory network for image restoration.
\newblock In: Proceedings of the IEEE international conference on computer
  vision. (2017)  4539--4547

\bibitem{jain2008natural}
Jain, V., Seung, S.:
\newblock Natural image denoising with convolutional networks.
\newblock Advances in neural information processing systems \textbf{21} (2008)

\bibitem{zhou2020awgn}
Zhou, Y., Jiao, J., Huang, H., Wang, Y., Wang, J., Shi, H., Huang, T.:
\newblock When awgn-based denoiser meets real noises.
\newblock In: Proceedings of the AAAI Conference on Artificial Intelligence.
  Volume~34. (2020)  13074--13081

\bibitem{cheng2021nbnet}
Cheng, S., Wang, Y., Huang, H., Liu, D., Fan, H., Liu, S.:
\newblock Nbnet: Noise basis learning for image denoising with subspace
  projection.
\newblock In: Proceedings of the IEEE/CVF Conference on Computer Vision and
  Pattern Recognition. (2021)  4896--4906

\bibitem{xie2012image}
Xie, J., Xu, L., Chen, E.:
\newblock Image denoising and inpainting with deep neural networks.
\newblock Advances in neural information processing systems \textbf{25} (2012)

\bibitem{ulyanov2018deep}
Ulyanov, D., Vedaldi, A., Lempitsky, V.:
\newblock Deep image prior.
\newblock In: Proceedings of the IEEE conference on computer vision and pattern
  recognition. (2018)  9446--9454

\bibitem{ren2018dn}
Ren, H., El-Khamy, M., Lee, J.:
\newblock Dn-resnet: Efficient deep residual network for image denoising.
\newblock In: Asian Conference on Computer Vision, Springer (2018)  215--230

\bibitem{zhang2017beyond}
Zhang, K., Zuo, W., Chen, Y., Meng, D., Zhang, L.:
\newblock Beyond a gaussian denoiser: Residual learning of deep cnn for image
  denoising.
\newblock IEEE transactions on image processing \textbf{26} (2017)  3142--3155

\bibitem{wang2017dilated}
Wang, T., Sun, M., Hu, K.:
\newblock Dilated deep residual network for image denoising.
\newblock In: 2017 IEEE 29th international conference on tools with artificial
  intelligence (ICTAI), IEEE (2017)  1272--1279

\bibitem{wang2019multi}
Wang, Y., Wang, G., Chen, C., Pan, Z.:
\newblock Multi-scale dilated convolution of convolutional neural network for
  image denoising.
\newblock Multimedia Tools and Applications \textbf{78} (2019)  19945--19960

\bibitem{mao2016image}
Mao, X., Shen, C., Yang, Y.B.:
\newblock Image restoration using very deep convolutional encoder-decoder
  networks with symmetric skip connections.
\newblock Advances in neural information processing systems \textbf{29} (2016)

\bibitem{liu2018denoising}
Liu, J.Y., Yang, Y.H.:
\newblock Denoising auto-encoder with recurrent skip connections and residual
  regression for music source separation.
\newblock In: 2018 17th IEEE International Conference on Machine Learning and
  Applications (ICMLA), IEEE (2018)  773--778

\bibitem{woo2018cbam}
Woo, S., Park, J., Lee, J.Y., Kweon, I.S.:
\newblock Cbam: Convolutional block attention module.
\newblock In: Proceedings of the European conference on computer vision (ECCV).
  (2018)  3--19

\bibitem{hu2018squeeze}
Hu, J., Shen, L., Sun, G.:
\newblock Squeeze-and-excitation networks.
\newblock In: Proceedings of the IEEE conference on computer vision and pattern
  recognition. (2018)  7132--7141

\bibitem{liu2020residual}
Liu, J., Zhang, W., Tang, Y., Tang, J., Wu, G.:
\newblock Residual feature aggregation network for image super-resolution.
\newblock In: Proceedings of the IEEE/CVF conference on computer vision and
  pattern recognition. (2020)  2359--2368

\bibitem{zhang2018ffdnet}
Zhang, K., Zuo, W., Zhang, L.:
\newblock Ffdnet: Toward a fast and flexible solution for cnn-based image
  denoising.
\newblock IEEE Transactions on Image Processing \textbf{27} (2018)  4608--4622

\bibitem{ren2021adaptive}
Ren, C., He, X., Wang, C., Zhao, Z.:
\newblock Adaptive consistency prior based deep network for image denoising.
\newblock In: Proceedings of the IEEE/CVF Conference on Computer Vision and
  Pattern Recognition. (2021)  8596--8606

\bibitem{mou2021dynamic}
Mou, C., Zhang, J., Wu, Z.:
\newblock Dynamic attentive graph learning for image restoration.
\newblock In: Proceedings of the IEEE/CVF International Conference on Computer
  Vision. (2021)  4328--4337

\bibitem{anwar2019real}
Anwar, S., Barnes, N.:
\newblock Real image denoising with feature attention.
\newblock In: Proceedings of the IEEE/CVF international conference on computer
  vision. (2019)  3155--3164

\bibitem{chang2020spatial}
Chang, M., Li, Q., Feng, H., Xu, Z.:
\newblock Spatial-adaptive network for single image denoising.
\newblock In: European Conference on Computer Vision, Springer (2020)  171--187

\bibitem{yue2019variational}
Yue, Z., Yong, H., Zhao, Q., Meng, D., Zhang, L.:
\newblock Variational denoising network: Toward blind noise modeling and
  removal.
\newblock Advances in neural information processing systems \textbf{32} (2019)

\bibitem{pang2021recorrupted}
Pang, T., Zheng, H., Quan, Y., Ji, H.:
\newblock Recorrupted-to-recorrupted: Unsupervised deep learning for image
  denoising.
\newblock In: Proceedings of the IEEE/CVF Conference on Computer Vision and
  Pattern Recognition. (2021)  2043--2052

\bibitem{huang2021neighbor2neighbor}
Huang, T., Li, S., Jia, X., Lu, H., Liu, J.:
\newblock Neighbor2neighbor: Self-supervised denoising from single noisy
  images.
\newblock In: Proceedings of the IEEE/CVF Conference on Computer Vision and
  Pattern Recognition. (2021)  14781--14790

\bibitem{zheng2021deep}
Zheng, H., Yong, H., Zhang, L.:
\newblock Deep convolutional dictionary learning for image denoising.
\newblock In: Proceedings of the IEEE/CVF Conference on Computer Vision and
  Pattern Recognition. (2021)  630--641

\bibitem{lai2018fast}
Lai, W.S., Huang, J.B., Ahuja, N., Yang, M.H.:
\newblock Fast and accurate image super-resolution with deep laplacian pyramid
  networks.
\newblock IEEE transactions on pattern analysis and machine intelligence
  \textbf{41} (2018)  2599--2613

\bibitem{da2014method}
Da, K.:
\newblock A method for stochastic optimization.
\newblock arXiv preprint arXiv:1412.6980 (2014)

\bibitem{paszke2019pytorch}
Paszke, A., Gross, S., Massa, F., Lerer, A., Bradbury, J., Chanan, G., Killeen,
  T., Lin, Z., Gimelshein, N., Antiga, L.,  et~al.:
\newblock Pytorch: An imperative style, high-performance deep learning library.
\newblock Advances in neural information processing systems \textbf{32} (2019)

\bibitem{aharon2006k}
Aharon, M., Elad, M., Bruckstein, A.:
\newblock K-svd: An algorithm for designing overcomplete dictionaries for
  sparse representation.
\newblock IEEE Transactions on signal processing \textbf{54} (2006)  4311--4322

\bibitem{gu2014weighted}
Gu, S., Zhang, L., Zuo, W., Feng, X.:
\newblock Weighted nuclear norm minimization with application to image
  denoising.
\newblock In: Proceedings of the IEEE conference on computer vision and pattern
  recognition. (2014)  2862--2869

\bibitem{guo2019toward}
Guo, S., Yan, Z., Zhang, K., Zuo, W., Zhang, L.:
\newblock Toward convolutional blind denoising of real photographs.
\newblock In: Proceedings of the IEEE/CVF conference on computer vision and
  pattern recognition. (2019)  1712--1722

\bibitem{kim2020transfer}
Kim, Y., Soh, J.W., Park, G.Y., Cho, N.I.:
\newblock Transfer learning from synthetic to real-noise denoising with
  adaptive instance normalization.
\newblock In: Proceedings of the IEEE/CVF Conference on Computer Vision and
  Pattern Recognition. (2020)  3482--3492

\bibitem{liu2021invertible}
Liu, Y., Qin, Z., Anwar, S., Ji, P., Kim, D., Caldwell, S., Gedeon, T.:
\newblock Invertible denoising network: A light solution for real noise
  removal.
\newblock In: Proceedings of the IEEE/CVF Conference on Computer Vision and
  Pattern Recognition. (2021)  13365--13374

\end{thebibliography}

\end{document}